\documentstyle[sprocl]{article}

\newtheorem{hardy}{Hardyism}
\newtheorem{khrennikov}{Khrennikovism}
\newtheorem{wootters}{Woottersism}
\newtheorem{preskill}{Preskillism}

\newcommand{\be}{\begin{equation}}
\newcommand{\ee}{\end{equation}}
\newcommand{\bea}{\begin{eqnarray}}
\newcommand{\eea}{\end{eqnarray}}
\newcommand{\bq}{\begin{quotation}}
\newcommand{\eq}{\end{quotation}}

\newcommand{\blh}{\begin{hardy}\protect$\!\!${\em\bf :}$\;\;$}
\newcommand{\elh}{\end{hardy}}
\newcommand{\bakh}{\begin{khrennikov}\protect$\!\!${\em\bf :}$\;\;$}
\newcommand{\eakh}{\end{khrennikov}}
\newcommand{\bbw}{\begin{wootters}\protect$\!\!${\em\bf :}$\;\;$}
\newcommand{\ebw}{\end{wootters}}
\newcommand{\bjp}{\begin{preskill}\protect$\!\!${\em\bf :}$\;\;$}
\newcommand{\ejp}{\end{preskill}}

\begin{document}

\title{The Anti-V\"axj\"o Interpretation of Quantum Mechanics}

\author{Christopher A. Fuchs \smallskip
\\
\it Computing Science Research Center
\\
\it Bell Labs, Lucent Technologies
\\
\it Room 2C-420, 600--700 Mountain Ave.
\\
\it Murray Hill, New Jersey 07974, USA}

\date{1 April 2002}

\maketitle

\bigskip

\begin{abstract}
In this note, I try to accomplish two things.  First, I fulfill
Andrei Khrennikov's request that I comment on his ``V\"axj\"o
Interpretation of Quantum Mechanics,'' contrasting it with my own
present view of the subject matter.  Second, I try to paint an image
of the hopeful vistas an information-based conception of quantum
mechanics indicates.
\end{abstract}
\smallskip

Andrei Khrennikov has asked me to make a few remarks contrasting my
view of the foundations of quantum mechanics to his view---something
he calls ``The V\"axj\"o Interpretation of Quantum
Mechanics.''~\footnote{See A. Khrennikov, ``V\"axj\"o Interpretation
of Quantum Mechanics,'' {\tt quant-ph/0202107}.} I would certainly be
loath to let him down on this occasion, not least of all because of
the time and loving devotion he spent in organizing our beautiful
2001 midsummer's meeting in V\"axj\"o. Indeed, those who know Andrei
Khrennikov know that he loves nothing better than a good fight!  What
better tribute might I offer him than a playful article titled ``The
Anti-V\"axj\"o Interpretation?''

Still, the constraint imposed by a timely publication of this
proceedings volume makes it hard for me to do the job properly.  In
the present note I first content myself to giving the reader a few
pointers to some of the literature that I think best captures my own
view. Following that, I rely on the mode of expression that flows so
freely from me when I write a personal letter---but never flows from
me when I write a proper paper!---to give a brief survey of what I
deem to be the big picture of my efforts. Here goes.

First the pointers.  Much of my own (forming) view of what is going
on in quantum mechanics is documented at my website:

\begin{center}
{\tt http://netlib.bell-labs.com/who/cafuchs/}
\end{center}

\noindent In particular, my most relevant (though wordy) document
there is the one titled ``Quantum States:\ What the Hell Are They?\
({\it The Post-V\"axj\"o Phase Transition})'' Another important
website (though some of the material there is not completely in line
with the way I see things) is the one of Carlton Caves:\smallskip
\begin{center}
{\tt http://info.phys.unm.edu/$\tilde{\;\;}\!$caves/}
\end{center}\smallskip

\noindent There, the most important file is the one titled
``Resource Material for Promoting the Bayesian View of Everything.''

Beyond that, let me recommend four other articles. The first two are
the most technically important for the enterprise I promote in my
other contribution to this volume:  Namely, to secure a transfer
from our present abstract, axiomatic formulation of quantum
mechanics to a more physically meaningful one.  I think some
elements in Lucien Hardy's papers {\it almost\/} carry us to the
brink of that.  In his work, I think the right emphasis is finally
being placed on the right mathematical structures.  The papers are:
\begin{enumerate}
\item
L.~Hardy, ``Quantum Theory From Five Reasonable Axioms,'' \\ {\tt
quant-ph/0101012}.
\item
L.~Hardy, ``Why Quantum Theory?,'' {\tt quant-ph/0111068}.
\end{enumerate}
For its pleasant explanation of the similarities of the ``measurement
problem'' in classical physics and quantum physics, I recommend,
\begin{enumerate}
\setcounter{enumi}{2}
\item R.~Duvenhage, ``The Nature of Information in Quantum
Mechanics,'' {\tt quant-ph/0203070}.
\end{enumerate}
Finally, Chris Timpson's undergraduate(!)\ thesis deserves note for
its emphasis on the proper way to think about {\it what\/} the
Church-Turing Thesis is an attempt to formalize:
\begin{enumerate}
\setcounter{enumi}{3}
\item
C.~G. Timpson, ``Information and the Turing Principle:\ Some
Philosophical Considerations,'' available at {\tt
http://users.ox.ac.uk/$\tilde{\;\;}\!$quee0776/.}
\end{enumerate}
Getting ideas like that straight, I now believe, form the better
part of the uphill slope we must climb to understand quantum
mechanics.

I should point out, however, that in all four of the above
references, I think significant improvements could be made by
adopting a sufficiently Bayesian stance toward the use and meaning
of probability.

Now let me transfer attention to a set of personal letters I include
in this contribution.  The first three were written to Andrei
Khrennikov himself, and have explicitly to do with the contrast
between his views and mine.  In a nutshell, we both seem to believe
that quantum mechanics has something to do with setting ``contexts''
and seeing what unrolls from that.  Where we seem to disagree,
however, is in what that has to do with ``probability.''

The final two letters---the actual heart of this paper---were written
to John Preskill and William Wootters, each in turn.  These address
various broader themes that I think best set the tone for my present
research efforts.

\section{28 June 2001, to Khrennikov, ``Context Dependent
Probability''}

\bakh
It was nice to meet you in V\"axj\"o and discuss fundamental
problems of quantum theory.  Unfortunetely, I have the impression
that my presentation on Contextual Probabilistic Interpretation of
quantum theory was not so clear for participants (conversations
during lunches and dinners). I try to present my views as short and
clear as possible.
\eakh

Thank you for valuing my opinion on your ideas; I am flattered.  So
I treated the problem in a conscientious manner:  I downloaded and
read three of your papers ({\tt quant-ph/0103065}, {\tt
quant-ph/0105059}, and {\tt quant-ph/ 0106073}).

I am indeed quite intrigued by the possibility that quantum
mechanics may be nothing more than a calculus for comparing
probabilities when the experimental context cannot be deleted from
the results it brings about.  In vague philosophical terms, I think
this is precisely the kind of idea Bohr, Heisenberg, and Pauli were
bandying about in constructing their interpretation of quantum
mechanics.  It is certainly the kind of notion Bohr was trying to
get at with his emphasis on ``complementarity.''  So I would welcome
a more precise way (a mathematical way) of expressing the essence of
all this.  I myself have been attracted to this sort of thing for a
long time:  it is a large part of the thread connecting my ``Notes on
a Paulian Idea''~\footnote{See C.~A. Fuchs, ``Notes on a Paulian
Idea:\  Foundational, Historical, Anecdotal \& Forward-Looking
Thoughts on the Quantum (Selected Correspondence),'' {\tt
quant-ph/0105039}.}---that is, that the observer sets the context,
and, in the words of Pauli, cannot be ``detached'' from what he
finds. Also you can find discussions of it in Sections 4 and 8 of
the large paper I was circulating at the conference, ``Quantum
Foundations in the Light of Quantum Information.''~\footnote{See
C.~A. Fuchs, ``Quantum Foundations in the Light of Quantum
Information,'' {\tt quant-ph/ 0106166}.} I say all this to make it
clear that I am more than sympathetic to your program.

However, as much as I would like to tell you otherwise (because you
are my friend), I do not see that your present formulation of the
problem moves very far toward quantum mechanics in a convincing
way.  There are problems on at least two levels.

Maybe the most devastating and immediate is your move between Eqs.
(5) and (6) of {\tt quant-ph/0106073}.  (I'll focus on that paper
for specificity since I did not see you make a stronger argument in
either of the other two papers.)  You write:
\bq
\noindent
The perturbation term $\delta({\cal S},{\cal S}^\prime)$ depends on
absolute magnitudes of probabilities.  It would be natural to
introduce normalized coefficient of the context transition \ldots\
\eq
\raggedbottom The question anyone will ask is, ``Why is this
natural?'' What compels the precise form of the normalization other
than that it forces the equation to look of a more quantum
mechanical form.  Why did you choose the square root rather than the
third root, say? Indeed, why not divide by the absolute value of
$\delta$, or the exponential of $\delta$, or any other combination
of functions one could pull out of a hat?  To put it not so gently,
it looks as if you built the desired answer in at the outset, with
little justification otherwise.

The second level of my problem is that, even if you do get this far,
how do you make the further step to vector space representations of
quantum mechanics?  Why are observables POVMs and not other exotic
entities?  What leads us to the starting point of Gleason's
theorem?  Etc., etc.?  I don't see that you have enough structure to
do that.  But more importantly, until you have done that I would
have to say that your theory remains fairly empty in making a
connection to quantum mechanics.  Too empty.

The way I view the problem presently is that, indeed, quantum theory
is a theory of contextual probabilities.  This much we agree on:
within each context, quantum probabilities are nothing more than
standard Kolmogorovian probabilities.  But the contexts are set by
the structure of the Positive Operator-Valued Measures:  one
experimental context, one POVM.  The glue that pastes the POVMs
together into a unified Hilbert space is Gleason's ``noncontextuality
assumption'': where two POVMs overlap, the probability assignments
for those outcomes must not depend upon the context.  Putting those
two ideas together, one derives the structure of the quantum state.
The quantum state (uniquely) specifies a {\it compendium\/} of
probabilities, one for each context.  And thus there are
transformation rules for deriving probabilities in one context from
another.  This has the flavor of your program.  But getting to that
starting point from more general considerations---as you would like
to do (I think)---is the challenge I haven't yet seen fulfilled.

I very much hope that I have not offended you with these comments. I
greatly respect your program.  But because of that I want much from
it.  I want it to stretch our understanding.  John Wheeler used to
say, ``We must make as many mistakes as we can, as fast as we can, or
we'll never have a hope of gaining a true understanding!''  I let
that philosophy rule my research life.  Thus I can only commend you
for your exploration, and hold the strongest hope that something firm
will come from it with a little more work and contemplation.

\section{4 July 2001, to Khrennikov, ``Invitation''}

\bakh
Yes, this is very well! However, for me, the only bridge between
``reality'' and our subjective description is given by relative
frequencies.
\eakh

But there other ways to make the bridge:  this is what gambling
situations (like the Dutch-book argument that R\"udiger Schack spoke
about)~\footnote{See C.~M. Caves, ``Betting Probabilities and the
Dutch Book,'' at {\tt http://info.phys.unm.
edu/$\tilde{\;\;}\!$caves/reports/reports.html}.} are about. They
give a {\it non}frequency {\it operational\/} definition to
probabilities. Subjective probabilities make their OBJECTIVE mark on
the world by specifying how an agent should act when confronted with
them.

\section{4 July 2001, to Khrennikov, ``Context Dependent Subjective
Probability''}

\bakh
P.S. But! How can you unify contextuality with subjective
probability?
\eakh

I just don't see this as a problem.  In choosing one experiment over
another, I choose one context over another.  The experiment elicits
the world to do something.  To say that the world is indeterministic
means simply that I cannot predict with certainty what it will do in
response to my action.  Instead, I say what I can in the form of a
probability assignment.  My probability assignment comes about from
the information available to me (how the system reacted in other
contexts, etc., etc.).  Similarly for you, even though your
information may not be the same as mine.  The OBJECTIVE content of
the probability assignment comes from the fact that {\it no one\/}
can make {\it tighter\/} predictions for the outcomes of experiments
than specified by the quantum mechanical laws.  Or to say it still
another way, it is the very existence of transformation {\it rules\/}
from one context to another that expresses an objective content for
the theory.  Those rules apply to me as well as to you, even though
our probability assignments {\it within\/} each context may be
completely different (because they are subjective).  But, if one of
us follows the proper transformation rules---the quantum rules---for
going to one context from another, while the other of us does not,
then one of us will be able to take advantage of the other in a
gambling match.  The one of us that ignores the structure of the
world will be bitten by it!

\section{18 February 2002, to Preskill, ``Psychology 101''}

Let me reply to some of your points in a way that doesn't reflect
their original order.

\bjp
In the past I have sensed that you and I differ in how we regard
ourselves. I believe that I am just another physical system governed
by the same fundamental laws as any other system. You seem to think
there is a fundamental distinction between yourself and the system
you are observing. To me the Everett view is appealing because it
turns away from this egocentrism.
\ejp

It's funny, but when I read this, my reaction went in two rather
peculiar directions.  First I thought, ``I wonder if, in the end, the
only thing the great quantum foundations struggles will leave behind
is a few psychological observations?  If so, what a shame.''  But
secondly, I imagined Galileo hoisting me up to the top of the
Leaning Tower of Pisa and dropping me off it along with his two
famous stones.  Even though I cursed and screamed the whole way
down, I went ``splat'' at the same time that they went ``thud.''

Here's the psychological thought in a little more detail.  One of
the things that bugs me about the Everett view is what {\it I\/}
consider {\it its\/}  extreme egocentrism!  Now, how can that
be---both of us accusing the other's view as {\it the\/} egocentric
view? I'll tell you what I think, trying to express the problem from
both sides of the fence.

My side gets to go first.  What I find egocentric about the Everett
point of view is the way it purports to be a means for us little
finite beings to get outside the universe and imagine what it is
doing as a whole.  And what is it doing as a whole?  Something
fantastic?  Something almost undreamable?!  Something inexpressible
in the words of man?!?!  Nope.  It's conforming to a scheme some guy
dreamed up in the 1950s.

This whole fantastic universe can be boiled down to something
representable within one of its most insignificant components---the
brain of man.  Even toying with that idea, strikes me as an
egocentrism beyond belief.  The universe makes use of no principle
that cannot already be stuffed into the head of an average PhD in
physics?  The chain of logic that leads to the truth of the
four-color theorem (apparently) can't be stuffed into our heads, but
the ultimate operating principle for all that ``is'' and ``can be''
can?

It's a funny thing:  I don't think I've met anyone who would imagine
that mathematics will ever come to an end.  Or even that it {\it
can\/}  come to an end.  There'll always be new axiom sets to play
with, new formal structures to write down.  But with physics it's a
completely different story.  People are always wanting to say, ``Well
we've finally gotten there.''  Or, ``Even though we're not there,
we're pretty damned close.''  It's OK, even condoned, to have Dreams
of a Final Theory.  From this point of view, all the mathematics yet
to come is worthless as far as the essence of the universe goes; the
wad was already shot.

You get the point.  It's a psychological one, but it's one that I
find overwhelmingly powerful.  It is that anytime any of us ever has
the chutzpah to say, ``Here's an ultimate statement about reality,''
or even a potentially ultimate one, what we're really doing is
painting the world in the image of man.  We're saying that the measly
concepts we've managed to develop up to this point in time fit the
world in a way that none of our previous concepts have, that none in
the future will ever do better, and, most importantly, we view this
not as a statement about ourselves and the situation set by our
present evolutionary and intellectual stage, but rather as a
property of the universe itself.

Now let me start moving toward the other side of the fence.  The
question someone like me---someone who has these kinds of
blasphemous thoughts---has to ask himself is, how can I ever hope to
be a scientist in spite of all this?  What can science and all the
great achievements it has given rise to in the last 400 years be
about if one chooses to suspend one's dreams of a final theory at
the very outset?  (Or, to tribute Johnny [Wheeler], how can one have
law without law?)

I think the solution is in nothing other than holding
firmly---absolutely firmly---to the belief that we, the scientific
agents, are physical systems in essence and composition no different
than much of the rest of the world.  But if we do hold firmly to
that---in a way that I do not see the Everettistas holding to
it---we have to recognize that what we're doing in the game of
science is swimming in the thick middle of things.  We're swimming
in this undulant sea, and doing our best to keep our heads above the
water:  All the concepts that arise in a physical theory must be
interpreted to do with points of view we can construct from {\it
within\/} the world.

That is to say, we have to loosen the idea that a physical law is a
mirror image of what ``is'' in the world, and replace it with
something that expresses instead how each of us can best cope with
and hope to take advantage of the world exterior to ourselves. This,
it seems to me, is something that by its very definition can be
stuffed into the human brain.  The current state of science is our
presently best known means for survival.  A scientific theory
indeed, from this point of view, is yet another expression of
Darwinian principles. Scientific theories evolve and survive because
the survivors have a kind of staying power that none of the rest of
the competition have.  Not because they are part of the blueprint of
the universe.

The situation of quantum mechanics---I become ever more
convinced---illustrates this immersion of the scientific agent in
the world more clearly than any physical theory contemplated to
date.  That is because it tells you you have to strain really hard
and strip away most of the theory's operational content, most of its
workaday usefulness, to make sense of it as a reflection of ``what
is'' (independent of the agent) and---importantly---you insist on
doing that for all the terms in the theory.

I know you're going to find the last sentence debatable, but that is
what I see as the danger in the Everett point of view:  You are
able---or at least purportedly so---to view the universal state as a
reflection of something, but at the cost of deleting all the
concrete things it was meant to reflect in the first place.  What I
mean by this is, if we take any concrete situation in quantum
mechanics---a system, a measuring device, and some kind of model for
the beginning stages of a measurement---we can indeed construct a
Church-of-the-Larger-Hilbert space description of it.  I'll grant
you that.  But try to go the other way around without any
foreknowledge of the ``measurement'':  Start with the Church, and try
to derive from it that a concrete measurement has taken place, and
you encounter an embarrassment of riches.  You don't know how to
identify the valid worlds, etc., etc.  (And, if you ask me, invoking
decoherence as a cure-all is little more than a statement of faith
that some guy in Los Alamos has the all the answers to all the tough
questions the rest of us are too lazy to work out.)

So, I myself am left with a view of quantum mechanics for which the
main terms in the theory---the quantum states---express nothing more
than the gambling commitments I'm willing to make at any moment.
When I encounter various other pieces of the world, if I am
rational---that is to say, Darwinian-optimal---I should use the
stimulations those pieces give me to reevaluate my commitments. This
is what quantum state change is about.  The REALITY of the world I
am dealing with is captured by two things in the present picture:
\begin{enumerate}
\item I posit systems with which I find myself having encounters, and
\item I am not able to see in a deterministic fashion the stimulations
(call them measurement outcomes, if you like) those systems will
give me---something comes into me from the outside that takes me by
surprise.
\end{enumerate}

OK, now let me put myself squarely in your pasture.  You worry that
having those main terms in the theory refer to {\it my\/} (or {\it
your}, or Joe Buck's)~\footnote{See pages 125 and 156 of {\tt
quant-ph/0105039}.} gambling commitments, is committing a kind of
egocentrism. What respectable theory would refer to my particular
vices, my desires, my bank account in making its most important
statements?

This is going to surprise you now, but I agree with you
wholeheartedly.  Even enthusiastically so.  Where I seem to disagree
is that I do not find this a good reason to promote those vices,
those commitments to an unearthly realm and call them ``states of the
universe'' (or relative states therein).  Instead, it seems to me to
be a call to recognize them for what they are and to redouble our
efforts for getting at the real nub of the matter.

Let me try to give you a way of thinking about this that you might
respect.  What was Einstein's greatest achievement in getting at
general relativity?  For the purposes of the present exposition, I
would say it was in his recognizing that the ``gravitational field''
one feels in an accelerating elevator is just a coordinate
effect---it is something that is induced purely with respect to the
description of an observer.  In this light, the program of trying to
develop general relativity thus boiled down to trying to recognize
all the things within gravitational and motional phenomena that
should be viewed as consequences of our coordinate choices.  Or to
use a phrase I've come to like, it was in identifying all the things
that can be viewed as ``numerically additional'' to the observer-free
situation which come about purely by bringing the observer
(scientific agent, coordinate system, etc.) onto the scene.

Now the point is, that was a really useful process.  For in weeding
out all the things that can be viewed as ``merely'' coordinate
effects, the fruit left behind could be seen in a clear view for the
first time:  It was the Riemannian manifold that we call spacetime.

What I dream for in my foundational program for quantum mechanics is
something just about like that.  Weed out all the terms that have to
do with gambling commitments (I used to call it information,
knowledge, or belief), and what is left behind will play a role much
like Einstein's manifold.

This much of the program, I hope and suspect you will understand
even if you are not sympathetic to it.  But, I don't know, you might
be sympathetic to it.  (Especially if I've done a good job above.)
However, it is also true that you have rightly suspected some
tendencies in me that go further.  In particular, in opposition to
the picture of general relativity, where reintroducing the
coordinate system---i.e., reintroducing the observer---changes
nothing about the manifold (it only tells us what kind of sensations
the observer will pick up), I do not suspect the same of the quantum
world.  This is why I recommend to all my friends that they read
William James's little article ``The Sentiment of
Rationality.''~\footnote{W.~James, ``The Sentiment of Rationality,''
in {\sl The Writings of William James:\ A Comprehensive Edition},
edited with an introduction by J.~J. McDermott (Modern Library,
Random House, New York, 1967), pp.~317--345.} It sort of sets the
right mindset, even though it has nothing to do with quantum
mechanics (other than in the efficacy of taking gambles) and goes
much further on religion than I myself would go.

Anyway, here I suspect that reintroducing the observer will be more
like introducing matter into pure spacetime, rather than simply
gridding it off with a coordinate system.  ``Matter tells spacetime
how to curve {\it when it is there}, and spacetime tells matter how
to move {\it when it is there}.''  Observers, scientific agents, a
necessary part of reality?  No.  But do they tend to change things
once they are on the scene.  Yes.  Or at least that's the idea.

Does that mean that the scientific agent is something outside of
physical law?  Well, to give this an answer, you've got to go back
and be very careful to use the picture of ``physical law'' that I
built up at the beginning of the essay.  What we are ``governed'' by,
God only knows.  He's the one, if anyone, who sits outside the
physical universe and has a chance to look back at it whenever he
pleases. Our task is to build up as good and solid a set of beliefs
as we can from within it.  In that way, we increase our survival
power, and use our spare time to try to bring forth a few progeny of
our own. (I used the word ``governed,'' by the way, because you had
used it above.)

If Galileo had dropped me from the tower, I feel pretty confident
that I would have gone splat.

\section{25 February 2002, to Wootters, ``A Wonderful Life''}

Thanks for the two notes, and wow, thanks for reading the James
essay.  Your questions were anything but naive.  In fact, they were
much needed.  In trying to answer them, I think I significantly
clarified---to myself even!---what I'm hoping to get at.  Besides, I
certainly don't have a final stand yet; the whole point of view is
in the process of formation and questions like yours really help.

I'll do my best to reply to your questions below, and in the process
I think I'll finally compose what I've been wanting to say about
your ``private-world-within-entanglement'' musings.  At the end of
the note, I'll list some of the open questions on my mind.  (These
are likely to be the naive ones!)

\bbw
Of course I'm very sympathetic to the perspective you express in this
paragraph \ldots\ but couldn't one still argue that as a matter of
methodology, the tactic of pretending that we can know the whole
story has served science well?  We make up a model of the world, and
this model gives us something to shoot at.  We hang on to the model
until we have found an explicit flaw in it (other than the flaw of
hubris).  And then we move on to a new model.

I find this an interesting question.  On the one hand, I think this
strategy does work well in advancing science.  On the other hand,
scientists (and others) are much too prone to accept as true the
pragmatic lie that says we can fully understand the world.

Your note to John [Preskill] goes some way toward laying out an
alternative methodology.  You speak of science in Darwinian terms:
the most successful theories survive.  How then do we proceed as
scientists? I suppose the answer is that we still make up theories
and test them, but the theories are not tentative descriptions of
the world. Rather, theories are schemes for making predictions.  But
you obviously also want to say that our theories tell us {\em
\underline{something}} about reality, even if they are not
descriptions of reality. Moreover, our theories will tell us more
about reality if we identify and remove from them those aspects that
are subjective.  So your view of science is not entirely operational.
There is realism in the background.

Have I understood you correctly?
\ebw

Yes there is certainly a kind of realism working in the back of my
mind, if what you mean by ``realism'' is that one can imagine a world
which never gives rise to man or sentience of any kind.  This, from
my view, would be a world without science, for there would be no
scientific agents theorizing within it.  This is what I mean by
realism:  That man is not a priori the be-all and end-all of the
world.  (The qualification ``a priori'' is important and I'll come
back to it later.)

A quick consequence of this view is that I believe I eschew all
forms of idealism.  Instead, I would say all our evidence for the
reality of the world comes from without us, i.e., not from within
us.  We do not hold evidence for an independent world by holding
some kind of transcendental knowledge.  Nor do we hold it from the
practical and technological successes of our past and present
conceptions of the world's essence.  It is just the opposite.  We
believe in a world external to ourselves precisely because we find
ourselves getting unpredictable kicks (from the world) all the
time.  If we could predict everything to the final T as Laplace had
wanted us to, it seems to me, we might as well be living a dream.

To maybe put it in an overly poetic and not completely accurate way,
the reality of the world is not in what we capture with our
theories, but rather in all the stuff we don't.  To make this
concrete, take quantum mechanics and consider setting up all the
equipment necessary to prepare a system in a state $\Pi$ and to
measure some noncommuting observable $H$.  (In a sense, all that
equipment is just an extension of ourselves and not so very
different in character from a prosthetic hand.)  Which eigenstate of
$H$ we will end up getting as our outcome, we cannot say.  We can
draw up some subjective probabilities for the occurrence of the
various possibilities, but that's as far as we can go.  (Or at least
that's what quantum mechanics tells us.)  Thus, I would say, in such
a quantum measurement we touch the reality of the world in the most
essential of ways.

With that said, I now want to be very careful to distance this
conception of reality, from what I'm seeking in the foundation game
of quantum mechanics.  Here's the way I originally put it to John
[Preskill] the other day.  Let me repeat a good bit of it so that
it's at the top of your mind:

\bq
OK, now let me put myself squarely in your pasture.  You worry that
having those main terms in the theory refer to {\it my\/} (or {\it
your}, or Joe Buck's) gambling commitments, is committing a kind of
egocentrism. What respectable theory would refer to my particular
vices, my desires, my bank account in making its most important
statements?

This is going to surprise you now, but I agree with you
wholeheartedly.  Even enthusiastically so.  Where I seem to disagree
is that I do not find this a good reason to promote those vices,
those commitments to an unearthly realm and call them ``states of the
universe'' (or relative states therein).  Instead, it seems to me to
be a call to recognize them for what they are and to redouble our
efforts for getting at the real nub of the matter.  \ldots

What I dream for in my foundational program for quantum mechanics is
something just about like that.  Weed out all the terms that have to
do with gambling commitments (I used to call it information,
knowledge, or belief), and what is left behind will play a role much
like Einstein's manifold.

This much of the program, I hope and suspect you will understand
even if you are not sympathetic to it.  \ldots\  However, it is also
true that you have rightly suspected some tendencies in me that go
further.  In particular, in opposition to the picture of general
relativity, where reintroducing the coordinate system---i.e.,
reintroducing the observer---changes nothing about the manifold (it
only tells us what kind of sensations the observer will pick up), I
do not suspect the same of the quantum world.  \ldots

Anyway, here I suspect that reintroducing the observer will be more
like introducing matter into pure spacetime, rather than simply
gridding it off with a coordinate system.  ``Matter tells spacetime
how to curve {\it when it is there}, and spacetime tells matter how
to move {\it when it is there}.''  Observers, scientific agents, a
necessary part of reality?  No.  But do they tend to change things
once they are on the scene.  Yes.  Or at least that's the idea.
\eq

From some of my choices of words, I think you probably got the
impression that this thing---this structure within quantum
mechanics---that I'm hoping to find at the end of the day is meant
to be a model of ``reality.''  Or at least our ``current best guess''
of what reality is.  But no, that's not really what I want.  And your
questions helped make that much clearer to me.  Remember, for me,
the mark of reality is its indescribability.

What I'm asking for instead is something like what one finds in the
old movie, {\sl It's a Wonderful Life}.  That is to say, in our
scientific theories, we codify some fraction of what we know about
manipulating the world and conditionally predicting the phenomena
about us.  However, suppose we wanted to get at a measure of our
place in the world.  How would we quantify it, or at least qualify
it?  That is, how might we ask how important our lives and agential
actions are with respect to the theory we ourselves laid out?

Our only tool, of course, is the theory; for it defines the frame
for optimal thinking (and imagination) at any given moment.  We can
only gauge our measure by deleting the free variable that is
ourselves and seeing what is left behind.  You surely remember what
George Bailey found when his guardian angel granted his wish in {\sl
It's a Wonderful Life}.  He found that his life mattered.  So too is
what I suspect we will find in quantum mechanics.

But all of that is the sort of thing I won't be able to say in a
conference presentation for quite some time.  It's the sort of thing
that we discussed once before, in the context of some Jamesian
quote.  \footnote{In particular I was thinking about this quote of
William James:
\bq
The history of philosophy is to a great extant that of a certain
clash of human temperaments.  Undignified as such a treatment may
seem to some of my colleagues, I shall have to take account of this
clash and explain a good many of the divergencies of philosophies by
it.  Of whatever temperament a professional philosopher is, he
tries, when philosophizing, to sink the fact of his temperament.
Temperament is no conventionally recognized reason, so he urges
impersonal reasons only for his conclusions.  Yet his temperament
really gives him a stronger bias than any of his more strictly
objective premises.  It loads the evidence for him one way or the
other, making a more sentimental or more hard-hearted view of the
universe, just as this fact or that principle would.  He {\it
trusts\/} his temperament.  Wanting a universe that suits it, he
believes in any representation of the universe that does suit it. He
feels men of opposite temper to be out of key with the world's
character, and in his heart considers them incompetent and `not in
it,' in the philosophic business, even though they may far excel him
in dialectical ability.

Yet in the forum he can make no claim, on the bare ground of his
temperament, to superior discernment or authority.  There arises
thus a certain insincerity in our philosophic discussions:  the
potentest of all our premises is never mentioned.  I am sure it
would contribute to clearness if in these lectures we should break
this rule and mention it, and I accordingly feel free to do so.
\eq
} It's the underground reason for the philosophy.

At the level of convincing our peers, let me put it to you this
way.  Within quantum mechanics, there is an invariant piece which is
common to all of us by the very fact of our accepting the theory.
That is what we are in search of because in some sense---which need
not pertain to a realistic conception of a theory's correspondence
to nature---it is the core of the theory.  It is the single part
that we agree upon, even when we agree upon nothing else.  In the
direction I am seeking to explore, the quantum state is ``numerically
additional'' to that core.  (That is, the quantum state is a
compendium of Bayesian ``beliefs'' or ``gambling commitments'' and is
thus susceptible to the type of analysis James gives in his
``Sentiment of Rationality.''  Our particular choice of a quantum
state is something extra that we carry into the world.)

I hope that clears up some of the mystery of my thoughts for
you---it did for me.  Given John's implicit acceptance of the idea
that ``a true theory is a mirror image of nature,'' I should not have
said in my note that I agreed with him ``wholeheartedly.''  I do not
intend for {\it any\/} part of the formal structure of quantum
mechanics to be a mirror image of nature (in the sense of a proposed
final theory).  However, I do not intend to give up the reality of
our world either.

From my point of view, the only ``true'' reality that creeps into
quantum mechanics is ``in the differential''---i.e., in the changes
we induce upon our (personal) quantum states for this and that due to
any stimuli we give to or take from the outside world.  That,
however, is a pretty amorphous thing as theoretical entities go.  It
is little more than what might have been called in older language,
the measurement ``click.''

There is a temptation to go further---to say that the POVM element
$E_b$ associated with a measurement outcome $b$ is itself an element
of reality.  But I think that has to be resisted at all costs. There
are several arguments one can use to show that the {\it ascription\/}
of a particular POVM to a measurement phenomenon is a subjective
judgment at the same level of subjectivity as the quantum state
itself.  (In fact the two go hand in hand, one cannot support the
subjectivity of the quantum state without also taking the
subjectivity of the POVM.)  Instead, one should view the
(theoretical) ascription of a POVM to an actual measurement device
as an attempt to set the significance and meaning of the ``click'' it
elicits.  Similarly for the Krausian quantum operation associated
with the measurement:  It describes the subjective judgment we use
for updating our quantum-state assignment in the light of the
``click.''  (If you want more details about these arguments, I can
forward you some of my old write-ups on the subject.)

So, you probably ask by now, ``What does that leave for the core of
the theory?  Aren't you throwing away absolutely everything?''  And
the answer is, ``No, I don't think so.''  Let me give you an example
of something which I think is left behind.  Recall my favorite
argument for why the quantum state cannot be an element of
reality---it's the Einstein argument I wrote about in Section 3 of
my NATO paper.  Once I posit a state for a bipartite system, even
though by my own admission my actions are purely local, a
measurement on one of the systems can toggle the quantum state of
the other to a large range of possibilities.  Thus, I say that the
quantum state of the far-away system cannot be more than my
information or the compendium of subjective judgments I'm willing to
ascribe to that system.

Notice, however, that in positing the original state, I had to also
implicitly posit a tensor-product space for the bipartite system.
Let me ask you this:  Once this tensor-product space is set, is
there any way to toggle one of the factors from afar just as with
the quantum state?  As far as I can tell there is not.  Thus I would
say that the Hilbert space of the far-away system is a candidate for
part of the theory's core.  Well, the Hilbert space---once the
choice of a particular quantum state within it is excluded---really
carries no substance beyond its dimensionality $d$.  Thus, in a more
refined way of speaking, what I really mean to say is that when I
posit a quantum system, I am allowed to also posit a characteristic
property of it.  It is a property that can be captured by a single
integer $d$.

There are some other things which I can argue will be ``left behind''
in such an analysis, but I don't want to clutter this note too
much.  Mainly I presented the example above so that I could give you
a clearer sense of how I want to draw a distinction between the
rawest forms of ``reality'' (the surprises the world gives us) and
the ``core of a theory.''

It is the core of the theory (along with the theory as a whole) that
I am starting to view in Darwinian terms.  But don't we have every
right to posit that core as a property of the world itself, at least
as long as that belief serves us well?  This, as you point out, has
been the predominant image of what science is about heretofore.

The only answer I can give you is ``yes, we can'' (just as indeed we
have heretofore).  So, your point is well-founded.  What I am
worried about is whether we {\it should\/} posit it
so.\footnote{Here is the way I put it to Henry Folse when he asked,
``Every attempt to sketch a conception of the universe from our best
theories at any date in human history in effect commits such
arrogance. Were the Newtonians of the end of the seventeenth century
being ``egocentric'' to think that Sir Isaac had done nothing less
than peer into the mind of the Divine and discerned God's blueprints
for the universe?'':
\bq\noindent
Yes.  (In my opinion.)  And you might interpret James and pragmatism
in general as a reaction to that.  However, I think in our modern
age with quantum mechanics we have a motivation and opportunity in
front of us that James did not have.  Try to give quantum mechanics
a naive realist interpretation---you can do it, or at least both
Everett and Bohm tell us we can---and you find yourself contorting
yourself beyond belief.  It's as if nature is telling us for the
first time, ``Please don't interpret me in a naive realist fashion. I
can't stop you, but please don't.''
\eq
Folse, by the way, points out that what I really mean here is
Cartesian representational realism---that reality is as we represent
it mathematically in our theories---rather than naive realism.} You
say that this view has guided science well in the past. But how do
you know? In a world with a view that there is no ultimate law, how
do you know that we would not be a thousand years more advanced if
we had only better appreciated our role as the substratum of our
theories?  I think it boils down to the difference between an active
and passive view of what existence is about.  Or maybe the
difference between a positive and a negative view.

To make this point, let me try to put things back into the context
of regular Darwinian evolution.  Consider the word ``elephant.''
Does it denote anything that exists in a kind of timeless sense, in
a way that we usually think---or in my case, previously thought---of
physical theories as existing?  If the concept of an elephant is
worthy of treating as a candidate for an element of reality, then so
too will a theory's core.

Well, if we have bought into Darwinism in any serious way, then I
would say, no, there is nothing particularly timeless about the
concept of an elephant.  There was once a chance that it might not
even arise in the world.  The ``elephant'' is merely a function of
the selective pressures that cropped up in our world's particular
history.  And, ashes to ashes, dust to dust, the poor elephant may
eventually disappear from the face of the universe, just like so
many species that arose in the course of evolution only to be never
discovered by a single archeologist.

But now, contrast the evolution of the elephant with the possible
future evolution of the human species.  The elephant was an accident
pure and simple, from the strictly Darwinian view.  But I would be
hard pressed to apply pure Darwinism to the future of mankind.  The
birth of my oldest daughter, for instance was no accident.  Her
traits were selected based on personal visions that both her mother
and I had for the future.  Similarly, but not so excitingly, with
the golden retriever, and all our other domesticated species.  The
key point is that in the present stage of evolutionary development,
we have it within our power to move beyond strict Darwinism.  This
is what our industry of genetic engineering is all about.

However, we would have never gotten to this stage if we had not
first realized that the concept of a species is not immutable.  As
strange---and as crazy and as scary---as it may sound, this is where
my thoughts are starting to roam with physical theories.  This does
not mean, however, that we can have exactly what we want with our
physical theories---that they themselves are little more than
dreams.  Just as the genetic engineer can make a million viruses
that will never have a chance of surviving on their own, there is
more to the story than our whims and fancies:  There is the
ever-present selective pressure from the outside.  But that does not
delete the genetic engineer's ability to make something that was
never here before.

But now, I go far, far, far beyond what I needed to say to answer
all your questions.  Mainly, I just wanted to emphasize why I
intentionally placed the words ``a priori'' in my definition of
reality way above.

I fear now slightly that you're going to realize I'm one of the
craziest people you've ever met!  And, trust me, I'm not sure I
really believe all that I said in the last three paragraphs.  But it
does strike me as a productive, or at least hopeful, train of
thought that someone ought to explore.  I guess I offer myself as
the sacrifice.

\noindent --- --- --- --- ---

There. I think that's enough of my going around your questions in a
rather wide way.  Let me now zoom back to the center of one of them
for purposes of a final emphasis.

\bbw
But you obviously also want to say that our theories tell us {\em
\underline{something}} about reality, even if they are not
descriptions of reality.
\ebw\smallskip

I hope you can glean from all the above that I do indeed believe our
theories tell us something about reality.  But that something is
much like what the elephant tells us about reality.  It's presence
tells us something about the accumulated selective pressures that
have arisen up to the present date.  A theory to some extent is a
statement of history.  It is also a statement of our limitations
with respect to all the pressures yet seen, or---more carefully---a
statement of our limitations with respect to our imaginations for
classifying all that we've yet seen.  (I for instance, cannot jump
off the leaning tower of Pisa unprotected and hope to live; you, for
instance, cannot get into your car and hope to push on the
accelerator until you are traveling beyond the speed of light.)
Finally, to the extent that we the theory users are part of nature,
the theory also tells us something about nature in that way.

But for any theory, there is always something outside of it.  Or at
least that's the idea I'm trying to build.  \medskip

\noindent PS.  Way above, I said I would finally say a few words about
your ``private-world-within-entangle\-ment'' musings.  But somehow it
didn't quite fit in with the flow of the rest of what I wanted to
say.  So, let me try to present the statement in isolation.  From my
point of view, the quantum state, and with it entanglement, never
pierces into the quantum system for which we posit a parameter $d$
(the ``dimension'').  Similarly for any bipartite system for which we
posit two parameters $d_1$ and $d_2$.  The quantum state is only
about what I'm willing to bet will be the consequences when I reach
out and touch a system.  Otherwise, indeed, a quantum system denotes
a private world unto itself.  And similarly with bipartite systems.
We have very little right to say much of anything about the
goings-on of their insides.  (This part of the picture is something
I've held firmly for a long time; it even shows up in my {\it Physics
Today\/} article with Asher Peres.)   \medskip

\noindent PPS.  I also promised to end with some open questions.  But
I'm petered out now.  And if you've gotten this far, you're probably
exhausted too.  So I'll just leave it for the future, depending upon
how interesting you find the ideas above, or how much you think
they're nonsense!

\section{Acknowledgements}

I thank Jeff Bub, Henry Folse, Bas van Fraassen, Julio
Gea-Banacloche, Lucien Hardy, David Mermin, John Preskill, Terry
Rudolph, and Bill Wootters for saying things that encouraged
me---whether they wanted them to or not---to make these letters more
public than usual.

\end{document}